\documentclass[journal]{IEEEtran}
\usepackage{amsmath, amssymb, amsthm, mathtools, bbm, relsize, paralist, hyperref, cite}
\usepackage[T1]{fontenc}

\newtheorem{theorem}{Theorem}
\newtheorem{lemma}[theorem]{Lemma}

\newtheorem{definition}{Definition}
\newtheorem{example}{Example}
\newtheorem{remark}{Remark}

\newcommand{ \calX }{ {\mathcal X} }
\newcommand{ \calY }{ {\mathcal Y} }
\newcommand{ \calS }{ {\mathcal S} }
\newcommand{ \calU }{ {\mathcal U} }
\newcommand{ \calM }{ {\mathcal M} }
\newcommand{ \calC }{ {\mathcal C} }
\newcommand{ \prob }{ \mathbb{P} }

\newcommand{ \Yhat }{ \widehat{Y} }
\newcommand{ \What }{ \widehat{W} }
\newcommand{ \fhat }{ \widehat{f} }
\newcommand{ \ghat }{ \widehat{g} }
\newcommand{ \tauhat }{ \widehat{\tau} }
\newcommand{ \f }{ \textnormal{f} }
\newcommand{ \fl }{ \textsc{fl} }
\newcommand{ \vl }{ \textsc{vl} }
\newcommand{ \bl }{ \textsc{bl} }
\newcommand{ \vanish }{ \mathsmaller{\downarrow} }
\newcommand{ \si }{ \mathrm{si} }
\newcommand{ \SI }{ \mathrm{SI} }
\newcommand{ \scaus }{ \mathrm{sc} }
\newcommand{ \caus }{ \mathrm{c} }
\newcommand{ \ncaus }{ \mathrm{nc} }
\newcommand{ \nost }{ \textnormal{-} }
\newcommand{ \ack }{ \textsc{ack} }
\newcommand{ \nack }{ \textsc{nack} }
\newcommand{ \myqed }{ \hfill $\blacktriangle$ }
\newcommand{ \defeq }{ \coloneqq }

\begin{document}

\title{Fundamental Limits of Communication Over State-Dependent Channels With Feedback}

\author{
				Mladen~Kova\v{c}evi\'c, 
				Carol~Wang, and
				Vincent~Y.~F.~Tan 
\thanks{Date: January 15, 2019.
        
				This work was supported by the Singapore Ministry of Education (Grant no. R-263-000-C83-112).
				M. Kova\v{c}evi\'{c} was also partially supported by the European Commission (H2020 Antares project, ref. no. 739570).
				The work was presented at the 2018 IEEE International Symposium on Information Theory (ISIT) \cite{isit}.
				
				M. Kova\v{c}evi\'{c} was with the Department of Electrical \& Computer Engineering,
        National University of Singapore, Singapore 117583.
				He is now with the BioSense Institute,
				University of Novi Sad, 21000 Novi Sad, Serbia
				(email: kmladen@uns.ac.rs).
				
				C. Wang was with the Department of Electrical \& Computer Engineering,
        National University of Singapore, Singapore 117583.
				She is now an independent researcher
				(email: cwang.ecc@gmail.com).
				
        V. Y. F. Tan is with the Department of Electrical \& Computer Engineering,
        National University of Singapore, Singapore 117583,
				and the Department of Mathematics,
				National University of Singapore, Singapore 119076
				(email: vtan@nus.edu.sg).}%
}%


\maketitle

\begin{abstract}
The fundamental limits of communication over state-dependent discrete memoryless
channels with noiseless feedback are studied, under the assumption that the
communicating parties are allowed to use variable-length coding schemes.
Various cases are analyzed, with the employed coding schemes having either bounded
or unbounded codeword lengths, and with state information revealed to the encoder
and/or decoder in a strictly causal, causal, or non-causal manner.
In each of these settings, necessary and sufficient conditions for positivity of
the zero-error capacity are obtained and it is shown that, whenever the zero-error
capacity is positive, it equals the conventional vanishing-error capacity.
Moreover, it is shown that the vanishing-error capacity of state-dependent channels
is not increased by the use of feedback and variable-length coding.
Both these kinds of capacities of state-dependent channels with feedback are thus
fully characterized.%
\end{abstract}


\section{Introduction}
\label{sec:intro}

\IEEEPARstart{W}{ith} the aim of enhancing their performance and simplifying
the employed protocols, modern communication systems are often designed in a
way that one or both of the communicating parties have access to side information
about the channel through which the data is being exchanged.
Additionally, a feedback link is also frequently implemented, providing a means
for the transmitter to obtain information about the channel output and adapt
its further transmission accordingly.
The gains in performance obtained by exploiting both these kinds of additional
information are especially significant if one is able to use variable-length
transmission strategies in the particular scenario of interest, and to tolerate
random decoding delays that come as a consequence.
An example of a widely used communication scheme that is adaptive in nature
and in which the length of transmission depends on the information obtained
through the feedback link is the well-known ARQ mechanism.

For the above reasons, studying the fundamental limits of communication achievable
by variable-length coding schemes is an important problem, especially so in
channels with feedback and with side information available at the transmitter,
the receiver, or both.
In the present work, we address this problem for channels with states, which
have a long history in communications and information theory \cite{elgamal+kim, proakis}
and serve as models for time-varying communication environments (wireless
fading channels, channels with jammers, etc.), as well as for several other
data transmission and storage scenarios (write-once memories with programmed
cells, memories with stuck-at faults, etc.).
In particular, we derive the capacity of state-dependent discrete memoryless
channels with feedback under variable-length coding, and under different models
of state information availability.
In addition to the conventional capacity, which allows a small (asymptotically
vanishing) probability of error at the decoder, we also analyze---and this is in
fact a major part of this work---the \emph{zero-error} capacity, which is defined
under a more stringent requirement that the decoder's output must always be correct.

Our interest in error-free communication is partly motivated by the recent surge
of activity in the area of ultra-reliable communications for 5G wireless systems,
i.e., applications that call for transmission reliability of nearly 100\% \cite{durisi}.
While the gap between `nearly 100\%' and `exactly 100\%' is impossible to bridge
in the majority of standard communication channels without feedback, including
the binary symmetric channel (BSC) and the relatively benign binary erasure channel
(BEC), it became evident after Burnashev's work on the error exponent of discrete
memoryless channels (DMCs) \cite{burnashev} that, with \emph{variable-length encoding}
and noiseless \emph{feedback}, error-free communication is not only possible over
a large class of DMCs, but the zero-error capacity of such channels is equal to
their (vanishing-error) Shannon capacity.
In the present work we generalize this result to channels with random states
and feedback by showing that, for a large class of such channels, the channel
capacity can be achieved with zero-error coding schemes, i.e., schemes of ultimate
reliability, with essentially no loss in performance.

In the following subsection we describe in precise terms the channel model
that is assumed throughout the paper and define the basic notions to be used
later in the text.
Our main contributions are stated in Section \ref{sec:outline} together with
an outline of the remainder of the paper.

\subsection{The Channel Model, Definitions, and Notation}
\label{sec:definitions}

Let $ \calX, \calY, \calS $ denote the sets of channel input letters, channel
output letters, and channel states, respectively, all of which are assumed finite.
A state-dependent discrete memoryless channel (SD-DMC) is described by conditional
probability distributions $ W(y|x,s) $, $ x \in \calX $, $ y \in \calY $, $ s \in \calS $,
where the states are drawn i.i.d.\ across all channel uses according to a distribution
$ Q(\cdot) $ on $ \calS $.
To avoid discussing trivial cases, we assume throughout the paper that $ |\calX| \geq 2 $,
$ |\calY| \geq 2 $, $ |\calS| \geq 1 $;
that all states in $ \calS $ have positive probability:
\begin{equation}
\label{eq:statespos}
  \forall s \in \calS \quad Q(s) > 0 ;
\end{equation}
and that every channel output is reachable from at least one input in at least one
state:
\begin{equation}
\label{eq:everyy}
  \forall y \in \calY  \quad  \exists x \in \calX, s\in \calS  \quad  W(y|x,s) > 0 .
\end{equation}

We use the symbol $ \calM $ to denote the set of messages to be transmitted in a
particular communication setting.
The symbols $ M, X, Y, S $ denote random variables taking values in
$ \calM, \calX, \calY, \calS $, respectively, and the lower case versions $ m, x, y, s $
denote their realizations, i.e., elements of $ \calM, \calX, \calY, \calS $.
The random variable $ M $ representing the transmitted message is always assumed
to be uniform over $ \calM $.\linebreak
$ X^\infty $ is a shorthand for a random infinite sequence $ (X_1, X_2, \ldots) $,
$ X^n $ for a random finite sequence $ (X_1, \ldots, X_n) $, and
$ X_k^n $ for a subsequence $ (X_k, \ldots, X_n) $
(hence $ X_1^n \equiv X^n $).

We say that the encoder (resp. decoder) has causal state information if, before
the $ n $'th channel use, it can see all the past channel states as well as the
current---$ n $'th---state, i.e., it is  given the state sequence $ S^n $ and can
use it in the $ n $'th time slot for the encoding (resp. decoding) operation.
State information\linebreak is said to be strictly causal if only past states ($ S^{n-1} $)
are avail\-able at time instant $ n $, and it is said to be non-causal if all the
states ($ S^\infty $) are available at any time instant.
We consider the following cases of state information availability:
\begin{align}
\label{eq:states}
  &\SI \defeq  \\\nonumber
  &\Big\{ (\nost,  \nost), (\scaus, \nost), (\caus,  \nost), (\ncaus, \nost),
          (\scaus, \caus), (\caus,  \caus), (\ncaus, \caus), (\ncaus, \ncaus) \Big\} ,
\end{align}
where the first (resp. second) coordinate of $ \si \in \SI $ denotes state information
available at the encoder (resp. decoder) and -/sc/c/nc stand for none/strictly-causal/causal/non-causal.
The eight cases that have been omitted from $ \SI $ are discussed in Remark \ref{rem:states}
to follow.

\begin{definition}
\label{def:VLF}
\textnormal{
  Consider an SD-DMC with causal state information at both the encoder and the
decoder (i.e., $ \si = (\caus,  \caus) $).
An $ (\ell, |\calM|, \epsilon) $ \emph{variable-length feedback (VLF) code} for
the message set $ \calM $, where $ \ell $ is a positive real and $ 0 \leq \epsilon \leq 1 $,
is defined by:
\begin{itemize}
\item[1)]
A sequence of encoders $ f_n : \calM \times \calY^{n-1} \times \calS^n  \to  \calX $,
$ n \geq 1 $, defining channel inputs $ X_n = f_n(M, Y^{n-1}, S^n) $;
\item[2)]
A sequence of decoders $ g_n : \calY^n \times \calS^n  \to  \calM $, $ n \geq 1 $,
defining decoder's estimates of the transmitted message, $ g_n(Y^n, S^n) $;
\item[3)]
A positive integer-valued random variable $ \tau $ that depends on the channel
outputs $ Y^n $ and the channel states $ S^n $
and satisfies $ \mathbb{E}[\tau] \leq \ell $.
(Formally, $ \tau $ is a stopping time of the receiver filtration
$ \{ \sigma(Y^n, S^n) \}_{n=0}^\infty $.
It represents the code length in this setting.)
\end{itemize}
The decoder's final decision is computed at time $ \tau $, $ \widehat{M} = g_\tau(Y^\tau, S^\tau) $,
and it must satisfy $ \mathbb{P}\big[\widehat{M} \neq M\big] \leq \epsilon $.
}

\textnormal{
When $ \epsilon = 0 $, such a code is called a \emph{zero-error VLF code}.
}

\textnormal{
If there exists a constant $ b < \infty $ such that $ \tau \leq b $, such a code
is called a \emph{bounded-length feedback code}, and if $ \tau = b = \ell $, it is
called a \emph{fixed-length feedback code}.
}

\textnormal{
Definitions for the other cases of state information availability $ \si \in \SI $ are
the same except $ S^n $ in 1)--3) is replaced by $ S^0 $, $ S^{n-1} $, or $ S^\infty $
accordingly.
}
\hfill \myqed
\end{definition}

The rate of an $ (\ell, |\calM|, \epsilon) $ code is defined as
$ \frac{1}{\ell} \log_2 |\calM| $.
The vanishing-error capacity of a given channel is defined in the usual way as the
supremum of the code rates that are asymptotically achievable (as $ \ell \to \infty $)
with arbitrarily small error probability.
The zero-error capacity of a given channel is the supremum of the rates of all
zero-error codes for that channel \cite{shannon}.
Capacity is always denoted by $ C $, with subscripts and superscripts indicating
the channel and the coding schemes with respect to which it is defined as follows:
\begin{itemize}
\item
The first subscript is either ``$ 0 $'' or ``$ \downarrow $'' and serves to distinguish
between the zero-error and the vanishing-error case;
\item
The second subscript is either ``$ \f $'' or ``$ \nost $'' depending on whether or not
the feedback link is present;
\item
The third subscript is $ \vl $, $ \bl $, or $ \fl $, indicating that the capacity in
question is defined with respect to variable-length, bounded-length, or fixed-length
codes;
\item
Superscripts from the set $ \SI $ (see \eqref{eq:states})
are used to denote state information availability at the encoder and the decoder.
\end{itemize}
For example, $ C_{\vanish, \f,\vl}^{\ncaus,\nost} $ is the vanishing-error capacity
under variable-length feedback codes, where the encoder is given state information in
a non-causal manner and the decoder is given no state information;
$ C_{0,\nost,\bl}^{\caus, \caus} $ is the zero-error capacity under bounded-length coding
without feedback, and with state information revealed both to the encoder and to the
decoder in a causal manner; etc.

\begin{remark}
\label{rem:states}
\textnormal{
To conclude this section we explain briefly why, of the sixteen possible cases in
$ \{\nost,\scaus,\caus,\ncaus\}^2 $, only the cases in $ \SI $ in \eqref{eq:states}
are being considered.
}

\textnormal{
First, leaving out the four cases with strictly causal state information at the decoder,
$ \{\nost,\scaus,\caus,\ncaus\} \times \{\scaus\} $, is not a loss in generality.
This is because strictly causal state information at the decoder can be ``made causal''
by simply delaying the decoding process by one time slot.
Hence, from the viewpoint of capacity issues, $ (*, \scaus) $ is equivalent to
$ (*, \caus) $ for any $ * \in \{\nost, \scaus, \caus, \ncaus\} $.
}

\textnormal{
Second, of the four possible cases where the decoder has non-causal state information
($ \{\nost,\scaus,\caus,\ncaus\} \times \{\ncaus\} $) we only consider one---$ (\ncaus, \ncaus) $.
This is also not a loss in generality because knowing future states can be helpful to
the decoder only if the encoder also knows future states and is using them in the
encoding operation.
Otherwise, these states are independent of the channel inputs and no information can
be extracted from them.
Hence, for our purposes, $ (*,\ncaus) $ is equivalent to $ (*, \caus) $ for any
$ * \in \{\nost, \scaus, \caus\} $.
}

\textnormal{
Finally, note that $ (\nost, \caus) $ has not been included in $ \SI $ either.
This case is quite subtle and we have chosen to discuss it separately in Section \ref{sec:nocaus}.
The main issue here is that it is not clear how to define the code length, i.e., the
stopping time $ \tau $ (see Definition \ref{def:VLF}).
Namely, the decoder making a decision at time instant $ \tau' $ does not necessarily
mean that the transmission is over from the encoder's perspective.
This is because the decoder's decision is based on the channel outputs and the channel
states that it sees, and hence the encoder, not knowing the states, may actually never
realize that the decoding is completed and may continue transmitting.
As we shall see, this is especially important in the zero-error setting.
}
\myqed
\end{remark}

\subsection{Main Results and Paper Outline}
\label{sec:outline}

Section \ref{sec:vanishing} considers the vanishing-error VLF capacity of
state-dependent channels.
The main result presented here (Theorem~\ref{thm:vanishing}) is that feedback
and variable-length coding do not increase the capacity of SD-DMCs.%

In Section \ref{sec:zeroVL} we give a full characterization of the zero-error
VLF capacity of state-dependent channels (Theorem \ref{thm:positivityVL}) by
deriving necessary and sufficient conditions for positivity of this quantity
and by proving that, whenever it is positive, it equals the vanishing-error
capacity of the same channel.
We also discuss here the differences with respect to the fixed-length coding
scenario that was recently analyzed in \cite{bracher+lapidoth}.

Section \ref{sec:zeroBL} contains the corresponding characterization of the
zero-error feedback capacity under bounded-length coding (Theorem \ref{thm:positivityVL}).

Section \ref{sec:nocaus} is devoted to SD-DMCs with state information available
only at the decoder.
As discussed in Remark \ref{rem:states}, this model exhibits some peculiarities
in the analysis of variable-length coding schemes, and for that reason we discuss
it in a separate section.
The main results stated here are a sufficient condition for the zero-error
feedback capacity under variable-length coding to be positive, and a necessary
and sufficient condition for the zero-error feedback capacity under bounded-length
coding to be positive.

Section \ref{sec:conclusion} concludes the paper and lists several open problems
related to those discussed in the present work.


\section{Vanishing-Error Capacity}
\label{sec:vanishing}

We begin our study of SD-DMCs with the analysis of the vanishing-error capacity
and show that this fundamental limit remains unchanged even if the transmitter
has access to noiseless feedback and is allowed to use variable-length coding
schemes.
For SD-DMCs with state information available only at the transmitter,
the fact that feedback does not increase the capacity under \emph{fixed-length}
coding was shown in \cite{merhav+weissman}.

\begin{theorem}
\label{thm:vanishing}
For every $ \si \in \SI $,
$ C_{\vanish,\f,\vl}^{\si} = C_{\vanish,\f,\bl}^{\si} = C_{\vanish,\f,\fl}^{\si} =
  C_{\vanish,\nost,\fl}^{\si} $.
\end{theorem}
\begin{IEEEproof}
Deferred to the Appendix.
\end{IEEEproof}

\vspace{3mm}
Consequently, in the remainder of the paper we shall denote the vanishing-error
capacity (with or without feedback) simply by $ C_\vanish^{\si} $.

By Theorem \ref{thm:vanishing} and the known expressions for the vanishing-error
capacity of SD-DMCs under fixed-length coding \cite[Ch.~7]{elgamal+kim}, we get:
\begin{align}
\label{eq:cap1}
   C_{\vanish, \f,\vl}^{\nost,\nost}  &= C_{\vanish, \f,\vl}^{\scaus,\nost} =
   \max_{P_X} I(X;Y)   \\
\label{eq:cap2}
   C_{\vanish, \f,\vl}^{\caus,\nost}  &=
   \max_{P_U, f\, :\, \calU \times \calS \to \calX} I(U;Y)   \\
\label{eq:cap3}
   C_{\vanish, \f,\vl}^{\ncaus,\nost} &=
        \max_{P_{U|S}, f\, :\, \calU \times \calS \to \calX} \big( I(U;Y) - I(U;S) \big)   \\
\label{eq:cap4}
   C_{\vanish, \f,\vl}^{\scaus,\caus} &=
   \max_{P_X} I(X;Y|S)   \\
\label{eq:cap5}
   C_{\vanish, \f,\vl}^{\caus,\caus}  &=
   C_{\vanish, \f,\vl}^{\ncaus,\caus}  =
   C_{\vanish, \f,\vl}^{\ncaus,\ncaus} =
   \max_{P_{X|S}} I(X;Y|S) ,
\end{align}
where $ U $ denotes an auxiliary random variable with alphabet $ \calU $ of cardinality
$ |\calU| \leq |\calX| |\calS| $.


\section{Zero-Error Capacity: Variable-Length Codes}
\label{sec:zeroVL}

In this section we study the problem of \emph{error-free} communication over
SD-DMCs by using VLF coding schemes.

Communicating with zero error has been considered previously in many settings,
starting with Shannon's work \cite{shannon} on the zero-error capacity of DMCs
with and without feedback, and under fixed-length encoding.
We refer the reader to \cite{korner+orlitsky} for a review of this area.
The work most closely related to what we discuss here is that of Bracher--Lapidoth
\cite{bracher+lapidoth}, where the zero-error feedback capacity of state-dependent
channels was determined, under the assumptions that fixed-length encoding is being
used and that state information is available only at the encoder.
We also mention the works of Zhao--Permuter \cite{zhao+permuter}, where the
authors gave a characterization of the zero-error feedback capacity under fixed-length
encoding for channels with state information at both the encoder and the decoder
(but in which the state process is not necessarily memoryless and is even allowed
to depend on the channel inputs), and
Tallini--Al-Bassam--Bose \cite{tallini}, where zero-error VLF communication
over the binary {\sf Z}-channel was studied.

For a DMC $ W(\cdot|\cdot) $ satisfying the ``non-triviality'' assumption
\eqref{eq:everyy}, a necessary and sufficient condition for the zero-error VLF
capacity to be positive is:
\begin{equation}
\label{eq:dmcvl}
  \exists x \in \calX, y \in \calY  \quad  W(y | x) = 0 .
\end{equation}
Moreover, whenever this condition is satisfied we have $ C_{0,\f,\vl} = C_{\vanish} $.
Both facts follow from Burnashev's characterization of the error exponent of DMCs
under VLF coding schemes \cite{burnashev}---the corresponding error exponent is
infinite at all rates below $ C_{\vanish} $ if and only if \eqref{eq:dmcvl} holds.
As elaborated by Massey \cite{massey}, there is a simpler and more direct way of
proving these statements which can be extended to channels for which error exponents
are not known;
we will use a generalization of that argument in the proof that follows.
We will also use the following terminology from \cite{massey}: if $ W(y | x) = 0 $,
then $ y $ is said to be a ``disprover'' for $ x $, indicating the fact that
such an output $ y $ disproves the possibility that $ x $ was the corresponding
input.

\begin{theorem}
\label{thm:positivityVL}
Necessary and sufficient conditions for positivity of the zero-error VLF capacity of
SD-DMCs are as follows:
\begin{itemize}
\item[(a)]
For $ \si \in \{(\nost,\nost), (\scaus,\nost)\} $,  $ C_{0,\f,\vl}^{\si} > 0 $
if and only if
\begin{equation}
\label{eq:vl_pos1}
   \exists x \in \calX, y \in \calY  \quad  \forall s \in \calS  \quad  W(y | x, s) = 0 .
\end{equation}
\item[(b)]
For $ \si \in \{(\caus,\nost), (\ncaus,\nost)\} $,  $ C_{0,\f,\vl}^{\si} > 0 $
if and only if
\begin{equation}
\label{eq:vl_pos2}
   \exists y \in \calY  \quad  \forall s \in \calS  \quad  \exists x \in \calX  \quad  W(y | x, s) = 0 .
\end{equation}
\item[(c)]
For $ \si \in \{(\scaus,\caus), (\caus,\caus), (\ncaus,\caus), (\ncaus,\ncaus)\} $,
$ C_{0,\f,\vl}^{\si} > 0 $\linebreak if and only if
\begin{align}
\nonumber
   \exists  x, x' \in \calX,\, &y \in \calY, s \in \calS  \\
\label{eq:vl_pos3}
	 &W(y | x, s) = 0 \; \wedge \; W(y | x', s) > 0 .
\end{align}
\end{itemize}
\noindent
Moreover, for every $ \si \in \SI $, if $ C_{0,\f,\vl}^{\si} > 0 $ then
$ C_{0,\f,\vl}^{\si} = C_\vanish^{\si} $.
\end{theorem}
\begin{IEEEproof}
\begin{inparaenum}
\item[{\em (a)}]
In the case $ \si = (\nost, \nost) $ when neither side has any state information,
the channel is equivalent to the DMC
$ \widetilde{W}(y|x) \defeq \sum_{s\in{\mathcal S}} Q(s) W(y | x, s) $.
The condition \eqref{eq:vl_pos1} is just the condition for positivity of the
zero-error VLF capacity of this DMC (see \eqref{eq:dmcvl} and \eqref{eq:statespos})
and is therefore necessary and sufficient for $ C_{0,\f,\vl}^{\nost,\nost} > 0 $.

Note that \eqref{eq:vl_pos1} is then sufficient for $ \si = (\scaus, \nost) $
as well because this model of state information availability is stronger than
$ (\nost, \nost) $ and $ C_{0,\f,\vl}^{\scaus,\nost}  \geq  C_{0,\f,\vl}^{\nost, \nost} $.
To see that \eqref{eq:vl_pos1} is also necessary when $ \si = (\scaus, \nost) $,
suppose that this condition is not satisfied, meaning that for every input-output
pair $ (x, y) \in \calX \times \calY $ there exists a state $ s_{x,y} $ with
$ W(y|x,s_{x,y}) > 0 $.
Then every output sequence $ y_1 \cdots y_n $ can be produced by every input sequence
$ x_1 \cdots x_n $ with positive probability (if the state sequence happens to be
$ s_{x_1,y_1} \cdots s_{x_n,y_n} $).
This means that the decoder cannot decide with certainty at any point in time what the
transmitted message was, and therefore, zero-error communication in a finite average
number of channel uses is impossible.

\item[{\em (b)}]
Let $ \si = (\caus, \nost) $.
To prove the sufficiency of \eqref{eq:vl_pos2}, we use the so-called Shannon
strategy \cite{shannon2, elgamal+kim} which reduces a given SD-DMC $ W $ with
causal state information at the transmitter to a related DMC $ W' $ with input
alphabet $ \calU \defeq \calX^\calS $ (the set of all functions from $ \calS $
to $ \calX $), output alphabet $ \calY $, and $ W'(y|u) \defeq \sum_{s\in\calS} Q(s) W(y|u(s),s) $.
If a code is specified over the alphabet $ \calU $, one can use it to communicate
over the original SD-DMC $ W $ as follows:
if the channel state in the current---$ n $'th---time slot is $ s $, and the $ n $'th
symbol of the codeword is $ u $, then the transmitter sends $ x = u(s) $.
In other words, in this approach one treats the SD-DMC $ W $ with $ \si = (\caus, \nost) $
as the DMC $ W' $, and, in particular, any zero-error code for $ W' $ is by using
this method translated to a zero-error code for $ W $.
Therefore, to show that \eqref{eq:vl_pos2} implies that the zero-error VLF capacity
of $ W $ is positive, it is enough to show that it implies that the zero-error VLF
capacity of $ W' $ is positive.
This is indeed the case because
\begin{align}
\nonumber
  &\exists y \in \calY  \quad  \forall s\in \calS  \quad \exists x \in \calX  \quad  W(y|x,s) = 0   \\
\label{eq:shstr2}
  \Leftrightarrow \quad
  &\exists y \in \calY  \quad  \exists u\in \calU  \quad  \forall s\in \calS  \quad  W(y|u(s),s) = 0   \\
\label{eq:shstr3}
	\Leftrightarrow \quad
  &\exists y \in \calY, u \in \calU  \quad  W'(y|u) = 0 ,
\end{align}
where \eqref{eq:shstr2} holds because $ u $ is a function from $ \calS $ to $ \calX $,
and \eqref{eq:shstr3} follows from the definition of $ W' $.
The expression in \eqref{eq:shstr3} is equivalent to saying that the zero-error
VLF capacity of $ W' $ is positive (see \eqref{eq:dmcvl}).

Again, since the model $ (\ncaus, \nost) $ is stronger than $ (\caus, \nost) $, implying
that $ C_{0,\f,\vl}^{\ncaus,\nost}  \geq  C_{0,\f,\vl}^{\caus, \nost} $, it is enough to
show the necessity of \eqref{eq:vl_pos2} for $ (\ncaus, \nost) $.
Suppose that \eqref{eq:vl_pos2} does not hold, i.e., for every output letter $ y $ there
exists a state $ s_y $ such that $ W(y|x,s_y) > 0 $ for all input letters $ x $.
Then for any output sequence $ y_1 \cdots y_n $ the state sequence $ s_{y_1} \cdots s_{y_n} $
produces $ y_1 \cdots y_n $ with positive probability on \emph{any} input $ x_1 \cdots x_n $.
This means that the decoder cannot be certain, at any time instant $ n $, what was the
transmitted message.%

\item[{\em (c)}]
For the case $ \si=(\scaus,\caus) $ one can use the standard technique of treating
the output $ Y $ and the state $ S $ as a joint output $ (Y, S) $ and notice that
an SD-DMC with $ \si=(\scaus,\caus) $ and noiseless feedback \emph{is equivalent}
to the DMC $ \overline{W}(y,s|x) \defeq Q(s) W(y|x,s) $ with noiseless feedback
($ \overline{W} $ has $ \calX $ as the input alphabet and $ \calY \times \calS $
as the output alphabet).
Namely, before the $ n $'th channel use the transmitter obtains $ Y^{n-1} $ through
feedback and $ S^{n-1} $ as side information, which is equivalent to saying that it
obtains the previous outputs of $ \overline{W} $, that is $ (Y,S)^{n-1} $, through
feedback.
The condition \eqref{eq:vl_pos3} is the condition for positivity of the zero-error
VLF capacity of the DMC $ \overline{W} $ (see \eqref{eq:dmcvl}) and is therefore
necessary and sufficient for $ C_{0,\f,\vl}^{\scaus,\caus} > 0 $.

That \eqref{eq:vl_pos3} is also necessary for $ \si \in \{(\caus,\caus), (\ncaus,\caus), (\ncaus,\ncaus)\} $
(or, indeed, for any $ \si $) is obvious, for if \eqref{eq:vl_pos3} is not satisfied,
then in every state an arbitrary output $ y $ is reachable from either all inputs,
or from none of them.
Clearly, any such state is useless for zero-error communication.%
\footnote{We should note that, for $ \si \in \{(\caus,\caus), (\ncaus,\caus), (\ncaus,\ncaus)\} $,
the necessity and sufficiency of \eqref{eq:vl_pos3} also follows from the results
in \cite{como}, where the error exponent of finite-state ergodic Markov channels
with causal state information at both sides was characterized.
Namely, the error exponent of an SD-DMC with causal state information at both sides
is infinite at all rates below capacity if and only if \eqref{eq:vl_pos3} holds.}
\end{inparaenum}

It is left to prove the final part of the statement claiming that the zero-error VLF
capacity, whenever positive, equals the vanishing-error capacity of the corresponding
channel.
Since $ C_{0,\f,\vl}^{\si}  \leq  C_{\vanish,\f,\vl}^{\si}  \equiv  C_\vanish^{\si} $,
only the achievability part needs to be shown, and this is done by using the Han--Sato
coding scheme.
This scheme was described in \cite{han+sato} for DMCs and bounded-length codes, but
virtually no changes are required to adapt it to the setting we are interested in,
so we only give an informal sketch of the argument here and refer the reader to
\cite{han+sato} for details.

Assume that $ C_{0,\f,\vl}^{\si} > 0 $.
Let $ \calM $ denote the set of all possible messages.
We first define two different codes for this message set:
\begin{inparaenum}
\item[1.)]
$ \calC $, a fixed-length feedback code of rate $ R \approx C_\vanish^{\si} $
and error probability less than $ \epsilon $ (we know that such a code exists for
any $ \epsilon > 0 $ and sufficiently large block-lengths), and
\item[2.)]
$ \calC_0 $, a zero-error VLF code of positive rate (which exists since $ C_{0,\f,\vl}^{\si} > 0 $).
\end{inparaenum}
Based on the codes $ \calC $ and $ \calC_0 $ we shall devise another
variable-length zero-error coding scheme of rate arbitrarily close to $ C_\vanish^{\si} $,
which will prove the desired claim.
The communication protocol is as follows.
To send a message $ m \in \calM $, the transmitter first sends the corresponding
codeword from $ \calC $.
Depending on whether or not the receiver has decoded the received sequence correctly,
something that the transmitter knows because it can simulate the decoding process after
receiving feedback, the transmitter then sends one bit of information through the channel.
This bit has the meaning of an $ \ack/\nack $ signal that informs the receiver about
the correctness of decoding, and can be transmitted \emph{error free} in a finite
expected number of channel uses because the zero-error VLF capacity is positive by
assumption.
Now, if the sent signal is $ \ack $, meaning that the decoding was correct and that
both the transmitter and the receiver are aware of that, the protocol stops.
If on the other hand the signal was $ \nack $, meaning that the decoding was incorrect,
the transmitter sends the same message again, but this time it encodes the message using
the zero-error code $ \calC_0 $, rather than the code $ \calC $.
This ensures that the receiver will decode the received sequence correctly with
probability $ 1 $ and the coding scheme just described is therefore zero-error.
Moreover, the overall rate of the scheme is approximately equal to the rate of the
code $ \calC $ used in the first phase of the protocol,
because the second phase of the protocol is active with probability at most $ \epsilon $
(the probability of incorrect decoding in the first phase), and this can be made
arbitrarily small.
\end{IEEEproof}

\begin{remark}
\textnormal{
Note that the condition \eqref{eq:vl_pos2} for positivity of the zero-error VLF
capacity is the same for causal and non-causal state information at the transmitter.
This is not the case in the fixed-length and bounded-length settings; see
\cite{bracher+lapidoth} and Theorem \ref{thm:positivityBL} ahead.
}

\textnormal{
Likewise, the condition \eqref{eq:vl_pos3} states that the zero-error VLF capacities
for the $ (\scaus,\caus) $ and $ (\caus, \caus) $ cases are either both positive or
both zero.
This is not the case when fixed-length or bounded-length codes are being used;
see Theorem~\ref{thm:positivityBL}.
\myqed
}
\end{remark}

We conclude this section with an example that is meant to demonstrate the power of
variable-length coding compared to fixed-length coding in channels with feedback---%
with fixed-length codes, information obtained by the transmitter through the feedback
link is not fully exploited.

\begin{example}[An SD-DMC with $ C_{0,\f,\fl}^{\ncaus,\ncaus} = 0 $ and
$ C_{0,\f,\vl}^{\nost,\nost} > 0 $]
\textnormal{
Consider the following binary-input-binary-output channel with two states:
in state $ s_0 $, we have the so-called {\sf Z}-channel with $ W(0|0,s_0) = 1 $ and
$ W(1|1,s_0) = 1 - p $, $ 0 < p < 1 $, and in state $ s_1 $ the channel is noiseless,
i.e., $ W(0|0,s_1) = W(1|1,s_1) = 1 $.
}

\textnormal{
Zero-error communication with fixed-length feedback codes through this channel is not
possible, even if both the transmitter and the receiver have non-causal state information.
This is because the state sequence may happen to be $ s_0 \cdots s_0 $ in which case
every two input sequences of length $ n $ are confusable, meaning that they can produce
the same output sequence with positive probability.
Hence, $ C_{0,\f,\fl}^{\ncaus,\ncaus} = 0 $.
}

\textnormal{
However, zero-error communication with variable-length feedback codes is possible even
if neither the transmitter nor the receiver have any state information, as one can
verify from \eqref{eq:vl_pos1} ($ y = 1 $ is a disprover for $ x = 0 $ in both states).
In fact, not only is it possible, but the zero-error VLF capacity is equal to the
vanishing-error capacity of the corresponding channel,
$ C_{0,\f,\vl}^{\nost,\nost} = C_{\vanish}^{\nost,\nost} $.
}
\myqed
\end{example}


\section{Zero-Error Capacity: Bounded-Length Codes}
\label{sec:zeroBL}

In the previous section we have demonstrated how variable-length encoding can
significantly increase the zero-error feedback capacity of an SD-DMC.
We now investigate the same problem in the situation where one wishes to impose a fixed
and deterministic upper bound on the codeword lengths, or equivalently on the stopping
time of transmission.
Variable-length codes in general have no such bound---even though their average length
is finite, each message is mapped to possibly infinitely many codewords of different
lengths, which means that the decoding delay can in general be arbitrarily large.
It is therefore natural, especially from the practical point of view, to consider
the case where the duration of transmission is upper bounded and to investigate the
corresponding fundamental limits.

Zero-error feedback capacity of DMCs under bounded-length coding
was first studied by Han and Sato \cite{han+sato}.
In particular, it was shown in \cite{han+sato} that the condition for positivity of
$ C_{0,\f,\bl} $ is the same as in the fixed-length case (with or without feedback)
\cite{shannon}, namely:
\begin{equation}
\label{eq:shannon}
  \exists x, x' \in \calX  \quad  \forall y \in \calY  \quad  W(y | x) W(y | x') = 0 .
\end{equation}
In words, the zero-error feedback capacity of a DMC under bounded-length (or fixed-length)
coding is positive if and only if there exist two non-confusable input letters.
However, the values of the two capacities are in general different:
while the zero-error feedback capacity under bounded-length coding is equal to the
vanishing-error capacity of the corresponding channel \cite{han+sato}, the zero-error
feedback capacity under fixed-length coding equals \cite{shannon}:
\begin{equation}
\label{eq:zedmc}
  \max_{P_X} \min_{y \in \calY}  - \log \sum_{x \in \calX : W(y | x) > 0} P_X(x) .
\end{equation}

As we show next, the situation is similar for state-dependent channels as well.%
\footnote{We note that, unlike for DMCs \cite{shannon}, the conditions
for positivity of the zero-error capacity under fixed-length coding for SD-DMCs
are in general not the same in the presence or absence of feedback.
E.g., there exist channels with $ C_{0,\f,\fl}^{\ncaus,\nost} > C_{0,\nost,\fl}^{\ncaus,\nost} = 0 $;
see \cite[Thm 7]{bracher+lapidoth}.
Hence, the situation is more subtle when channels have states, and not all results
are straightforward generalizations of their DMC counterparts.}

\begin{theorem}
\label{thm:positivityBL}
For every $ \si \in \SI $,
$ C_{0,\f,\bl}^{\si} > 0 $ if and only if $ C_{0,\f,\fl}^{\si} > 0 $.
In particular:
\begin{itemize}
\item[(a)]
For $ \si \in \{(\nost,\nost), (\scaus,\nost)\} $,  $ C_{0,\f,\bl}^{\si} > 0 $
if and only if
\begin{align}
\nonumber
  &\exists x, x' \in \calX  \quad  \forall y \in \calY  \\
\label{eq:bl_pos1}
  &\big( \forall s \in \calS  \quad  W(y | x, s) = 0 \big)   \vee
  \big( \forall s \in \calS  \quad  W(y | x', s) = 0 \big) .
\end{align}
\item[(b)]
$ C_{0,\f,\bl}^{\caus,\nost} > 0 $ if and only if there
exists a partition $ \calY_0, \calY_1 $ of $ \calY $ such that
\begin{equation}
\label{eq:bl_pos2}
  \forall s \in \calS  \quad  \exists x, x' \in \calX  \quad
  W(\calY_0 | x, s) = W(\calY_1 | x', s) = 1 .
\end{equation}
\item[(c)]
$ C_{0,\f,\bl}^{\ncaus,\nost} > 0 $ if and only if
\begin{align}
\nonumber
  \forall s, s' \in \calS  \quad  \exists x, x' \in\ &\calX  \quad  \forall y \in \calY  \\
\label{eq:bl_pos3}
  &W(y | x, s) W(y | x', s') = 0 .
\end{align}
\item[(d)]
$ C_{0,\f,\bl}^{\scaus,\caus} > 0 $ if and only if
\begin{equation}
\label{eq:bl_pos4}
  \exists x, x' \in \calX  \quad  \forall y \in \calY, s \in \calS  \quad
  W(y | x, s) W(y | x', s) = 0 .
\end{equation}
\item[(e)]
For $ \si \in \{(\caus,\caus), (\ncaus,\caus), (\ncaus,\ncaus)\} $,
$ C_{0,\f,\bl}^{\si} > 0 $ if and\linebreak only if
\begin{align}
\nonumber
  \forall s \in \calS  \quad  \exists x, x' \in \calX  \quad  &\forall y \in \calY  \\
\label{eq:bl_pos5}
  &W(y | x, s) W(y | x', s) = 0 .
\end{align}
\end{itemize}
Moreover, for every $ \si \in \SI $, if $ C_{0,\f,\bl}^{\si} > 0 $ then
$ C_{0,\f,\bl}^{\si} = C_{\vanish}^{\si} $.
\end{theorem}
\begin{IEEEproof}
We first show that fixed-length and bounded-length zero-error feedback capacities
are either both positive or both zero.
Since $ C_{0,\f,\bl}^{\si}  \geq  C_{0,\f,\fl}^{\si} $, the ``if direction'' is
trivial.
Conversely, suppose that $ C_{0,\f,\bl}^{\si}  >  0 $.
Then, for some $ 0 < \ell \leq n < \infty $, there exists a zero-error code of
cardinality at least $ 2 $, average length $ \ell $, and maximum length $ n $.
By ``zero-padding'' the codewords we can then construct a fixed-length zero-error
code of length $ n $ having the same cardinality, which implies that
$ C_{0,\f,\fl}^{\si}  \geq \frac{1}{n}  >  0 $.
Based on this observation, we can focus on fixed-length codes in proving the claims
\emph{(a)--(e)}.

\begin{inparaenum}
\item[{\em (a)--(c)}]
The conditions \eqref{eq:bl_pos1}--\eqref{eq:bl_pos3} for positivity of
$ C_{0,\f,\fl}^{\si} $ in cases when only the transmitter has state information
were derived in \cite[Thm 3, Thm 10, Rem. 17]{bracher+lapidoth}.
Note that \eqref{eq:bl_pos1} is the condition for positivity of the zero-error
fixed-length feedback capacity of the DMC
$ \widetilde{W}(y|x) = \sum_{s\in{\mathcal S}} Q(s) W(y | x, s) $; see \eqref{eq:shannon}.

\item[{\em (d)}]
As in the variable-length setting, the case $ \si=(\scaus,\caus) $ is solved by
observing that an SD-DMC with $ \si=(\scaus,\caus) $ and noiseless feedback is
equivalent to the DMC $ \overline{W}(y,s|x) \defeq Q(s) W(y|x,s) $ with noiseless
feedback.
The condition \eqref{eq:bl_pos4} is the condition for positivity of the zero-error
capacity of this DMC (see \eqref{eq:shannon}) and is therefore necessary and
sufficient for $ C_{0,\f,\fl}^{\scaus,\caus} > 0 $.

\item[{\em (e)}]
Let $ \si=(\caus,\caus) $.
If for every state $ s $ there exists a pair of non-confusable inputs $ x_s, x'_s $,
which is what the condition \eqref{eq:bl_pos5} means, then the transmitter and the
receiver can agree beforehand for $ x_s $ to mean $ 0 $ and $ x'_s $ to mean $ 1 $.
In this way, one bit can be transmitted error free in one channel use and so
$ C_{0,\f,\fl}^{\caus,\caus} > 0 $.

For the converse it is enough to consider the case $ \si=(\ncaus,\ncaus) $.
Suppose that \eqref{eq:bl_pos5} is not satisfied, meaning that there exists a
state $ s $ for which every two inputs are confusable.
If this is the case, then for the state sequence $ s^n = s \cdots s $ it is not
possible to transmit one bit error-free in any number of channel uses $ n $, and hence
$ C_{0,\f,\fl}^{\caus,\caus} = 0 $.
Informally, knowing the future states cannot help the encoder/decoder if these
states remain unfavorable throughout the entire transmission.

The proof of the fact that
$ C_{0,\f,\bl}^{\si} = C_{\vanish}^{\si} $ whenever $ C_{0,\f,\bl}^{\si} > 0 $
is analogous to the corresponding proof for the variable-length case (see Theorem
\ref{thm:positivityVL}).
\end{inparaenum}
\end{IEEEproof}


\section{State Information at the Decoder Only}
\label{sec:nocaus}

In this section we discuss SD-DMCs with state information available only at the decoder,
the case that has been left out of the discussion thus far.
As pointed out in Remark~\ref{rem:states}, in this channel model it is not quite clear
how to define the code length for variable-length codes, i.e., the stopping time of
the transmission (see Definition \ref{def:VLF}).
Since the decoder makes a decision based on the outputs $ Y^n $ and the states $ S^n $,
the encoder, not knowing the states, is not able to exactly simulate the decoding
process and to determine the moment when the decision has been made.
It can only provide an estimate of this moment based on the outputs $ Y^{n} $ which
it obtains through feedback.
As we shall see, this estimate is good enough when one considers coding with asymptotically
vanishing error probability (in fact, it is not even necessary as the vanishing-error
capacity can be achieved with fixed-length codes, with or without feedback).
However, in the case of zero-error communication the encoder has to be \emph{certain}
that the decoding was successful before it stops transmitting a given message and
starts transmitting the next message.
It is in this case that the effects of the mismatch in state information at the two
sides are most apparent.%

\begin{remark}
\textnormal{
Before proceeding with the analysis, we note that the channels studied in this
section (SD-DMCs with feedback and $ \si = (\nost, \caus) $) are in fact particular
instances of \emph{DMC's with noisy feedback}.
Namely, as already mentioned in the proofs of Theorems \ref{thm:positivityVL} and
\ref{thm:positivityBL}, the standard trick of treating the state information at
the decoder as part of a joint channel output $ (Y,S) $ of the DMC
$ \overline{W}(y,s|x) \defeq Q(s) W(y|x,s) $ shows that an SD-DMC $ W $ with
$ \si = (\scaus,\caus) $ and noiseless feedback is \emph{equivalent} to the DMC
$ \overline{W} $ with noiseless feedback (noiseless feedback in the DMC $ \overline{W} $
means that the transmitter sees $ (Y,S)^{n-1} $ before the $ n $'th channel use).
However, in the case $ \si = (\nost,\caus) $ this equivalence fails as the
transmitter now obtains only a degraded/noisy version ($ Y^{n-1} $) of the joint
output ($ (Y,S)^{n-1} $) through the feedback link.
}
\myqed	
\end{remark}

\subsection{Vanishing-Error Capacity}

We know from \cite[Ch. 7.4]{elgamal+kim} and \eqref{eq:cap4} that:
\begin{equation}
\label{eq:nocaus}
  C_{\vanish,\nost,\fl}^{\nost,\caus}  =  \max_{P_X} I(X;Y|S)  =  C_{\vanish,\f,\vl}^{\scaus,\caus} .
\end{equation}
Note that the issue with the stopping time mentioned above does not arise when
defining the two capacities in \eqref{eq:nocaus}:
for $ C_{\vanish,\nost,\fl}^{\nost,\caus} $ because in the fixed-length setting
the stopping time is fixed in advance, and for $ C_{\vanish,\f,\vl}^{\scaus,\caus} $
because in this case both sides have state information.

The quantity we are interested in here, $ C_{\vanish,\f,\vl}^{\nost,\caus} $,
has not been formally defined.
However, based on the obvious facts that the model $ (\scaus,\caus) $ is stronger
than $ (\nost,\caus) $, that having feedback is better than not having feedback,
and that fixed-length codes are a special case of variable-length codes, one
can still claim that the following chain of inequalities must hold:
$ C_{\vanish,\nost,\fl}^{\nost,\caus}
   \leq  C_{\vanish,\f,\bl}^{\nost,\caus}
   \leq  C_{\vanish,\f,\vl}^{\nost,\caus}
   \leq  C_{\vanish,\f,\vl}^{\scaus,\caus}  $.
From this and \eqref{eq:nocaus} we then conclude that
\begin{equation}
\label{eq:nocausvl}
  C_{\vanish,\f,\vl}^{\nost,\caus}  =  C_{\vanish,\f,\bl}^{\nost,\caus}  =  \max_{P_X} I(X;Y|S) = C_{\vanish}^{\scaus,\caus} .
\end{equation}
In particular, the VLF capacity of the $ (\nost,\caus) $ channel can be achieved by using
fixed-length codes.

\subsection{Zero-Error Capacity: Bounded-Length Codes}
\label{sec:nocbl}

We now turn to the zero-error problems and start with the bounded-length case.
The following theorem gives a necessary and sufficient condition for positivity
of the zero-error capacity in this setting.

\pagebreak
\begin{theorem}
\label{thm:nocausbl}
  The following statements are equivalent:
\begin{itemize}
\item[(a)]
$ C_{0,\f,\fl}^{\nost,\caus} > 0 $;
\item[(b)]
$ C_{0,\f,\bl}^{\nost,\caus} > 0 $;
\item[(c)]
$ C_{0,\f,\bl}^{\scaus,\caus} > 0 $;
\item[(d)]
\eqref{eq:bl_pos4} holds.
\end{itemize}
\end{theorem}
\begin{IEEEproof}
\begin{inparaenum}
\item[{\em (a) $ \Leftrightarrow $ (b)}:]
This is shown by ``zero-padding'' bounded-length codes to obtain fixed-length
codes, as for the other models of state information availability (see Theorem~\ref{thm:positivityBL}).

\item[{\em (b) $ \Rightarrow $ (c)}:]
This follows from $ C_{0,\f,\bl}^{\scaus,\caus} \geq C_{0,\f,\bl}^{\nost,\caus} $.

\item[{\em (c) $ \Leftrightarrow $ (d)}:]
This was shown in Theorem \ref{thm:positivityBL}(d).

\item[{\em (d) $ \Rightarrow $ (a)}:]
Suppose that \emph{(d)} holds, i.e., there exist two input letters $ x, x' $ that
are non-confusable in every state.
Then, if the transmitter sends $ x $ for $ 0 $ and $ x' $ for 1, the receiver will
be able to tell from the output which of the two possibilities is the correct one
because it knows the channel state.
Therefore, one bit can be transmitted in one channel use, and hence
$ {C_{0,\f,\fl}^{\nost,\caus} \geq 1 > 0} $.
\end{inparaenum}
\end{IEEEproof}

\vspace{3mm}
We now know from Theorem \ref{thm:nocausbl} that
$ C_{0,\f,\bl}^{\nost,\caus} > 0 \;\Leftrightarrow\; C_{0,\f,\bl}^{\scaus,\caus} > 0 $,
from Theorem \ref{thm:positivityBL} that
$ C_{0,\f,\bl}^{\scaus,\caus} = C_{\vanish}^{\scaus,\caus} $ whenever
$ C_{0,\f,\bl}^{\scaus,\caus} > 0 $, and from \eqref{eq:nocausvl} that
$ C_{\vanish}^{\scaus,\caus} = C_{\vanish}^{\nost,\caus} $.
It is then natural to ask if it is also true that
$ C_{0,\f,\bl}^{\nost,\caus} = C_{\vanish}^{\nost,\caus} $ whenever
$ C_{0,\f,\bl}^{\nost,\caus} > 0 $?
The corresponding equality for the models of state information availability from
$ \SI $ has been established in Theorem \ref{thm:positivityBL}, but the proof
technique used there does not apply when $ \si = (\nost,\caus) $.
The difficulty is precisely in the fact we mentioned at the beginning of this
section---in this model the transmitter cannot exactly simulate the decoding
process\linebreak
because it does not know the states.
Hence, the transmitter is in general not able to decide whether the receiver
has decoded the received sequence correctly, and whether it should retransmit
the same codeword or start sending the next codeword.
Therefore, for $ \si = (\nost,\caus) $ it is not clear whether the vanishing-error
capacity can be achieved with zero-error bounded-length codes whenever
$ C_{0,\f,\bl}^{\nost,\caus} > 0 $.
We next give an example of a channel for which the answer to this question is
affirmative.

\begin{example}
\label{ex:noc}
\textnormal{
Consider the following binary-input-binary-output channel with two states:
in state $ s_0 $ the channel flips the input bit, $ W(1|0,s_0) = W(0|1,s_0) = 1 $,
and in state $ s_1 $ it leaves the bit intact, $ W(0|0,s_1) = W(1|1,s_1) = 1 $.
}

\textnormal{
Suppose that the transmitter sends $ x $ in the $ n $'th slot and that $ y $ is
produced at the channel output.
After the transmitter obtains $ y $ through the feedback link, it can easily determine
the $ n $'th state: the state is $ s_1 $ if and only if $ x = y $.
This means that the transmitter effectively obtains (strictly causal) state information
through feedback and therefore
$ C_{0,\f,\bl}^{\nost,\caus} = C_{0,\f,\bl}^{\scaus,\caus} = C_{\vanish}^{\scaus,\caus} $,
where the second equality holds because $ C_{0,\f,\bl}^{\scaus,\caus} > 0 $
(see \eqref{eq:bl_pos4}).
\myqed
}
\end{example}

The main point in Example \ref{ex:noc} is the following: if the states are uniquely
determined by the channel inputs and outputs, then the problem of calculating
$ C_{0,\f,\bl}^{\nost,\caus} $ is reduced to the (easier) problem of calculating
$ C_{0,\f,\bl}^{\scaus,\caus} $, which was solved in Theorem \ref{thm:positivityBL}.
Whether such a reduction is possible in general is an interesting question that
we shall not be able to answer here.

\subsection{Zero-Error Capacity: Variable-Length Codes}
\label{sec:nocvl}

For variable-length codes, we can only give a sufficient condition for positivity
of the zero-error feedback capacity at this point.
The idea behind this condition is based on the fact that, for some channels, the
transmitter obtains state information ``for free'' through the feedback link, in
which case the $ (\nost,\caus) $ model is effectively reduced to the $ (\scaus,\caus) $
model (see Example \ref{ex:noc}).
As we show in Theorem \ref{thm:nocvl}, it is in fact not necessary that \emph{all}
states be uniquely determined by inputs and outputs in order for this reduction to
work---it is enough that only a group of states exists that contains a ``disprover''
and that is discernible from other states with positive probability.

\begin{theorem}
\label{thm:nocvl}
  The following condition is sufficient for $ {C_{0,\f,\vl}^{\nost,\caus} > 0} $:
\begin{align}
\nonumber
  \exists  &x, x'  \in \calX, y \in \calY, \calS^* \subseteq \calS, \calS^* \neq \emptyset  \\
\nonumber
	 &\big( \forall s \in \calS^*  \quad  W(y | x', s) > 0  \;\wedge\;  W(y | x, s) = 0 \big)  \;\;\wedge \quad  \\
\label{eq:nocvlpos}
    &\phantom{x}\big( \forall s \in \calS \setminus \calS^*  \quad  W(y | x', s) = 0 \big) .
\end{align}
\end{theorem}
\begin{IEEEproof}
  Let us first parse the condition \eqref{eq:nocvlpos}.
The meaning of the statement $ W(y | x', s) > 0  \;\wedge\;  W(y | x, s) = 0 $
is the same as before: $ y $ is a disprover for $ x $ (in the group of states $ \calS^* $).
Further, we require the existence of an event ($ x' \rightarrow y $) that has positive
probability in the group of states $ \calS^* $, but is impossible in other states.
The occurrence of this event serves to the transmitter as an identifier of the
group of states $ \calS^* $.

Now, assuming that \eqref{eq:nocvlpos} holds, one bit of information can be sent
with zero error as follows.
In the first two channel uses the transmitter sends $ x, x' $ for $ 0 $ and $ x', x $
for 1.
If the letters obtained at the output are $ \lnot y, y $ \emph{and} the state in
the second slot is from $ \calS^* $, then the receiver concludes that $ 0 $ must
have been sent.
The reason is the following: since the received symbol in the \emph{second} slot
is $ y $ and the state is from $ \calS^* $ (the receiver can see the states), the
transmitted symbol must have been $ x' $, and then it automatically follows that
the symbol sent in the first slot is $ x $.
Furthermore, the transmitter is also assured that the state in the second slot is
from $ \calS^* $ and that the receiver has received the bit correctly because the 
transition $ x' \rightarrow y $ is only possible in states from $ \calS^* $.
Similarly, if the letters obtained at the output are $ y, \lnot y $ \emph{and} the
state in the \emph{first} slot is from $ \calS^* $, then the receiver concludes that
$ 1 $ must have been sent, and the transmitter is assured that the receiver has received
the bit correctly.
In summary, if the channel output is either $ \lnot y, y $ or $ y, \lnot y $ and
the state in the slot in which the output is $ y $ is from $ \calS^* $, then the
protocol stops and both parties agree on the value of the transmitted bit.
If any other situation occurs, the procedure is repeated.
The expected number of steps needed to complete the transmission of the bit is
finite because the event that both parties are waiting for ($ x' $ produces $ y $
in a state from $ \calS^* $) has positive probability.
Therefore, $ C_{0,\f,\vl}^{\nost,\caus} > 0 $.
\end{IEEEproof}

\vspace{3mm}
Note that the condition \eqref{eq:vl_pos1} for $ C_{0,\f,\vl}^{\nost,\nost} > 0 $,
together with \eqref{eq:everyy}, implies the condition \eqref{eq:nocvlpos}
(as it should, because clearly $ C_{0,\f,\vl}^{\nost,\nost} > 0 $ implies
$ C_{0,\f,\vl}^{\nost,\caus} > 0 $).
To see that it does, suppose that \eqref{eq:vl_pos1} holds and take
$ \calS^* = \{s : W(y | x', s) > 0\} $.
The reverse implication does not hold because in \eqref{eq:vl_pos1} we require
that $ W(y | x, s) = 0 $ in all states, rather than just in states from $ \calS^* $
as in \eqref{eq:nocvlpos}.

Note also that the sufficient condition for $ C_{0,\f,\vl}^{\nost,\caus} > 0 $
given in \eqref{eq:nocvlpos} is different from the necessary and sufficient
condition for $ C_{0,\f,\vl}^{\scaus,\caus} > 0 $ given in \eqref{eq:vl_pos3}.
Based on Theorem \ref{thm:nocausbl}, one might wonder whether it holds that
$ C_{0,\f,\vl}^{\nost,\caus} > 0 \;\Leftrightarrow\; C_{0,\f,\vl}^{\scaus,\caus} > 0 $,
i.e., whether \eqref{eq:vl_pos3} is also necessary and sufficient for
$ C_{0,\f,\vl}^{\nost,\caus} > 0 $?
This is, however, not the case.
In the following example we describe a channel for which
$ C_{0,\f,\vl}^{\nost,\caus} = 0 $ and yet $ C_{0,\f,\vl}^{\scaus,\caus} > 0 $.
In other words, strictly causal state information at the transmitter can in
some cases enable the parties to communicate error free, even if they were
not able to do that in the absence of this information.
This is a somewhat curious fact having in mind that in most other settings
studied so far, strictly causal state information at the transmitter has been
shown equivalent (in the sense of achievable rates) to \emph{no} state information:
$ C_{\vanish}^{\nost,\nost}  =  C_{\vanish}^{\scaus,\nost} $,
$ C_{\vanish}^{\nost,\caus}  =  C_{\vanish}^{\scaus,\caus} $,
$ C_{0,\f,\bl}^{\nost,\nost}  =  C_{0,\f,\bl}^{\scaus,\nost} $,
$ C_{0,\f,\vl}^{\nost,\nost}  =  C_{0,\f,\vl}^{\scaus,\nost} $
(see also Theorem \ref{thm:nocausbl}).

\begin{example}[$ (\nost,\caus) \neq (\scaus,\caus) $]
\textnormal{
Consider the following binary-input-binary-output channel with two states:
in state $ s_0 $ the channel is a BSC($p$), where $ 0 < p < 1 $, i.e.,
$ W(0|0,s_0) = W(1|1,s_0) = 1 - p $, and in state $ s_1 $ the channel is noiseless,
$ W(0|0,s_1) = W(1|1,s_1) = 1 $.
}

\textnormal{
By Theorem \ref{thm:positivityVL}(c) we know that $ C_{0,\f,\vl}^{\scaus,\caus} > 0 $.
It is easy to see why this is the case---the transmitter and the receiver can agree on
the codeword termination by using the noiseless state since they both know the states
(the transmitter obtains the state information with a one-slot delay, but this can be
circumvented by sending a dummy letter).
}

\textnormal{
However, if the transmitter has no state information, i.e., if $ \si = (\nost,\caus) $,
then it is not possible to communicate with zero-error in a finite expected number of
channel uses.
To see this, suppose that the transmitter is trying to send one bit through the channel
by using a repetition code.
The receiver can see the channel states and will therefore recover the bit correctly
as soon as the state happens to be $ s_1 $, and it will be in a finite expected number
of channel uses.
However, the transmitter can never be sure whether this state has occurred and whether
it should stop transmitting.
This is because all the channel transitions that are possible in state $ s_1 $ are also
possible in state $ s_0 $ and therefore, based on the channel inputs (which it knows)
and the channel outputs (which it sees through feedback) alone, the transmitter cannot
determine with certainty whether any of the states so far was actually $ s_1 $.
In other words, if the transmitter wants to be sure that the bit was received correctly,
it must continue transmitting indefinitely.
(Note that the transmitter will occasionally recognize state $ s_0 $---if the channel input
is $ 0 $ and the output is $ 1 $, or vice versa---but this state is useless for zero-error
communication.)
It should be clear that the above conclusion holds for any coding scheme, not just for
repetition codes, and hence $ C_{0,\f,\vl}^{\nost,\caus} = 0 $.
}
\myqed
\end{example}


\section{Concluding Remarks and Further Work}
\label{sec:conclusion}

As we have seen, with fixed-length coding the information obtained by the transmitter
through the feedback link, as well as the side information about the channel states,
are not fully exploited.
For instance, by allowing the possibility of variable and adaptive transmission times
it is possible, under certain (mild) conditions, to achieve channel capacity with error probability
being fixed to zero.
Furthermore, for any fixed error probability $ \epsilon > 0 $, one can achieve higher
rates with variable-length coding compared to those achievable with fixed-length coding
(see the discussion in the Appendix).
For these reasons, variable-length coding schemes are a natural choice in systems with
feedback (whenever one is willing to tolerate random decoding delays), and it is therefore
important to study the corresponding fundamental limits of communication.

To conclude the paper, we state several problems, related to those we have analyzed here,
as pointers for further work:
\begin{itemize}
\item
Derive necessary and sufficient conditions for positivity of the zero-error VLF capacity
of SD-DMCs with state information available only at the decoder, $ C_{0,\f,\vl}^{\nost,\caus} $,
as well as the values of the capacities $ C_{0,\f,\bl}^{\nost,\caus} $ and
$ C_{0,\f,\vl}^{\nost,\caus} $ whenever they are positive
(see Sections~\ref{sec:nocbl} and \ref{sec:nocvl}).
\item
Investigate the corresponding questions about the zero-error VLF capacity when the
feedback is noisy, or incomplete, or when the receiver can also send coded information
over the feedback link (i.e., a function of the received sequence, rather than the
sequence itself).
See \cite{massey, asadi+devroye} for some results in this direction for DMCs.
\item
Investigate the corresponding questions in more general settings---multi-user
channels, channels with non-i.i.d. states, etc.
\item
Derive necessary and sufficient conditions for positivity of the zero-error capacity
of SD-DMCs \emph{without} feedback.
(This was solved in \cite{bracher+lapidoth} for $ \si = (\scaus,\nost) $ and $ \si = (\caus, \nost) $,
but the case $ \si = (\ncaus, \nost) $ was left open, see \cite[Thm 7]{bracher+lapidoth}.)
\end{itemize}

\section*{Acknowledgment}

We would like to thank Lan Vinh Truong (National University of Singapore), for
helpful discussions on VLF codes,
as well as the associate editor Giuseppe Durisi and the three anonymous referees,
for their thorough reviews and numerous suggestions for improvement.


\appendix
\section*{Proof of Theorem \ref{thm:vanishing}}

The theorem states that, for every $ \si \in \SI $,
$ C_{\vanish,\f,\vl}^{\si} = C_{\vanish,\f,\bl}^{\si} = C_{\vanish,\f,\fl}^{\si} =
  C_{\vanish,\nost,\fl}^{\si} $.
We prove this in two steps:
we first argue that $ C_{\vanish,\f,\bl}^{\si} = C_{\vanish,\f,\vl}^{\si} $,
and then we show in Lemma \ref{thm:epscap} that
$ (1-\epsilon) C_{\vanish,\f,\bl}^{\si} \leq C_{\vanish,\nost,\fl}^{\si} $,
$ \forall \epsilon > 0 $.
This, together with the obvious fact that
$ C_{\vanish,\f,\bl}^{\si}  \geq  C_{\vanish,\f,\fl}^{\si}  \geq  C_{\vanish,\nost,\fl}^{\si} $,
will conclude the proof.

To see that $ C_{\vanish,\f,\bl}^\si = C_{\vanish,\f,\vl}^\si $,
consider an arbitrary $ (\ell, |\calM|, \epsilon) $ variable-length code $ \calC_\vl $
whose codeword lengths are described by the stopping time $ \tau $ satisfying
$ \mathbb{E}[\tau] \leq \ell $.
One can then define a bounded-length code $ \calC_\bl $ which uses the same encoding
and decoding procedures as $ \calC_\vl $, except that the decoder is forced to make a
decision at time $ b $, if it hasn't already done so.
The error probability of the resulting code $ \calC_\bl $ is at most
$ \epsilon + \mathbb{P}[\tau > b]
  \leq \epsilon + \frac{\mathbb{E}[\tau]}{b}
  \leq \epsilon + \frac{\ell}{b} $,
where we have used Markov's inequality.
Since $ b $ can be taken arbitrarily large, this shows that there exists a bounded-length
code whose rate and error probability are arbitrarily close to the rate and error
probability of a given variable-length code.

We next prove that
$ (1-\epsilon) C_{\vanish,\f,\bl}^{\si} \leq C_{\vanish,\nost,\fl}^{\si} $,
for any $ {\epsilon > 0} $.
For concreteness, we consider the case $ \si = (\ncaus, \nost) $ (the Gel'fand--Pinsker
channel \cite{gelfand+pinsker}); the other cases can be obtained in a similar way.
The proof combines the approach from \cite{ppv} for DMCs with feedback, the derivation of
the capacity of the Gel'fand--Pinsker channel without feedback \cite[Sec. 7.6]{elgamal+kim},
and a certain inequality that is derived here and that is needed as a replacement for the
so-called Csisz\'{a}r sum identity.

\begin{lemma}
\label{thm:epscap}
For every $ \epsilon \in (0,1) $,
$ (1-\epsilon) C_{\vanish,\f,\bl}^{\si} \leq C_{\vanish,\nost,\fl}^{\si} $.
\end{lemma}
\begin{IEEEproof}
  Let $ \si = (\ncaus, \nost) $ and consider a particular bounded-length code $ (f_n, g_n, \tau) $
for this channel; see Definition~\ref{def:VLF}.
Following \cite[Thm 4]{ppv}, we define an extended channel with the input alphabet
$ \widehat{\mathcal X}   \defeq  {\mathcal X} \cup \{T\} $, the output alphabet
$ \widehat{\mathcal Y}   \defeq  {\mathcal Y} \cup \{T\} $, the set of states
$ \widehat{\mathcal S}   \defeq  {\mathcal S} $, and the transition probabilities
\begin{align}
  \What(\hat{y} | \hat{x}, \hat{s})  \defeq
	\begin{cases}
	  W(\hat{y} | \hat{x}, \hat{s}) , &  \hat{x} \neq T  \\
    1 ,      &  \hat{x} = T \land \hat{y} = T  \\
		0 ,      &  \hat{x} = T \land \hat{y} \neq T ,
	\end{cases}
\end{align}
as well as the corresponding code $ \big(\widehat{f}_n, \widehat{g}_n, \widehat{\tau}\big) $:
\begin{align}
\label{eq:vlft1}
  \fhat_n\big(M,\Yhat^{n-1},S^\infty\big)   &\defeq
	\begin{cases}
	  {f}_n\big(M,\Yhat^{n-1},S^\infty\big) , &  \tau \geq n  \\
    T ,                             &  \tau < n ,
	\end{cases}   \\
\label{eq:vlft2}
  \tauhat   &\defeq  \inf\big\{ n : \Yhat_n = T \big\}  =  \tau + 1   \\
  \ghat_n\big(\Yhat^{n}\big)   &\defeq
	\begin{cases}
	  {g}_n\big(\Yhat^{n}\big) ,              &  \tauhat >    n  \\
    {g}_n\big(\Yhat^{\tauhat - 1}\big) ,    &  \tauhat \leq n .
	\end{cases}
\end{align}
Here $ T \notin \calX \cup \calY $ is the ``termination'' symbol which is transmitted
noiselessly and serves for the transmitter to inform the receiver that the transmission
is over.
Apart from this addition, the two channels behave the same.
If $ (f_n, g_n, \tau) $ is an $ (\ell - 1, |{\mathcal M}|, \epsilon) $-bounded-length
code for the original channel satisfying $ \tau \leq b - 1 $, then
$ \big(\fhat_n, \ghat_n, \tauhat\big) $ is an $ (\ell, |{\mathcal M}|, \epsilon) $ code
for the extended channel with $ \tauhat \leq b $;
thus, upper bounding the rate of the former is essentially equivalent to upper bounding
the rate of the latter.

To derive the desired upper bound, recall that the Fano inequality asserts that
for any code $ \big(\fhat_n, \ghat_n, \tauhat\big) $ for the message set $ \mathcal M $
it holds that:
\begin{equation}
\label{eq:fano}
  (1 - \epsilon) \log |\calM| \leq I\big( M; \Yhat^b \big) + h(\epsilon) .
\end{equation}
In the following we show that the mutual information term in \eqref{eq:fano} can
be upper bounded as
$ I\big(M; {\Yhat}^b\big) \leq \ell \cdot C_{\vanish,\nost,\fl}^{\ncaus,\nost} + o(\ell) $.
Plugging this back into \eqref{eq:fano} would yield
$ \frac{1}{\ell} \log |\calM| \leq \frac{1}{1-\epsilon} C_{\vanish,\nost,\fl}^{\ncaus,\nost} + o(1) $
and would thus complete the proof of the lemma.

The following inequality was derived in the proof of \cite[Thm 4]{ppv}:
\begin{align}
\label{eq:app1}
  I\big(M; \Yhat^b\big)   \leq
		H(\tau)  + \sum_{k=1}^b I\big(M; \Yhat_k | V_k, \Yhat^{k-1}\big) ,
\end{align}
where
\begin{align}
\label{eq:Vn}
  V_n \defeq
	\begin{cases}
	  1 , &  \tauhat \leq n  \\
		0 , &  \tauhat >    n .
	\end{cases}
\end{align}
The derivation holds unchanged in our case too so we shall not repeat it here.
It was also shown there that
$ H(\tau) \leq (\ell + 1) h\big(\frac{1}{\ell + 1}\big) = o(\ell) $,
so we only need to upper bound the second summand on the right-hand side of \eqref{eq:app1}.
First notice that
\begin{align}
  \sum_{k=1}^b  I\big(M; \Yhat_k | V_k, \Yhat^{k-1}\big)
\label{eq:app2}
     &\leq \sum_{k=1}^b  I\big(M, \Yhat^{k-1}; \Yhat_k | V_k\big)  \\
\label{eq:app3}
     &=    \sum_{k=1}^b  \prob[V_k=0]\, I\big(M, Y^{k-1}; Y_k) ,
\end{align}
where \eqref{eq:app2} is by the chain rule for mutual information and
\eqref{eq:app3} is obtained by conditioning on the possible values of $ V_k $.
Namely, 
\begin{inparaenum}
\item[1.)]
conditioned on $ V_k = 1 $ we have $ \Yhat_k = T $ and hence the corresponding
mutual information term is zero, and
\item[2.)]
conditioned on $ V_k = 0 $, the statistics of the extended channel is identical
to that of the original channel and hence the mutual information term can be computed
for the latter.
\end{inparaenum}
Therefore,
$ I\big(M, \Yhat^{k-1}; \Yhat_k | V_k\big) =
  \prob[V_k=1] \cdot 0 + \prob[V_k=0] \cdot I\big(M, Y^{k-1}; Y_k) $.
We further have:
\begin{align}
\nonumber
  &\sum_{k=1}^b  \prob[V_k=0]\, I\big(M, Y^{k-1}; Y_k)   \\
\nonumber
		&\;=    \sum_{k=1}^b  \prob[V_k=0] 
		                 \Big(I\big(M, Y^{k-1}, S_{k+1}^b; Y_k \big)   \\
\label{eq:app31}
											    &\phantom{XXXXXXXXXX} - I\big(Y_k; S_{k+1}^b | M, Y^{k-1} \big) \Big)   \\
\nonumber
		&\;\leq \sum_{k=1}^b  \prob[V_k=0]
		                 \Big(I\big(M, Y^{k-1}, S_{k+1}^b; Y_k \big)   \\
\label{eq:app4}
													&\phantom{XXXXXXXXXX} - I\big(Y^{k-1}; S_k | M, S_{k+1}^b \big) \Big)   \\
\nonumber
		&\;=    \sum_{k=1}^b  \prob[V_k=0]
		                 \Big(I\big(M, Y^{k-1}, S_{k+1}^b; Y_k \big)   \\
\label{eq:app5}
													&\phantom{XXXXXXXXXX} - I\big(M, Y^{k-1}, S_{k+1}^b; S_k \big) \Big)   \\
\label{eq:app6}
		&\;=    \sum_{k=1}^b  \prob[V_k=0]
		                 \Big(  I\big(U_k; Y_k \big) - I\big(U_k; S_k \big) \Big)   \\
\label{eq:app8}
		&\;\leq \ell\, C_{\vanish,\nost,\fl}^{\ncaus,\nost}  ,
\end{align}
where \eqref{eq:app4} is shown below;
\eqref{eq:app5} holds because $ (M, S_{k+1}^b) $ is independent from $ S_k $;
in \eqref{eq:app6} we have denoted $ U_k = \big(M, Y^{k-1}, S_{k+1}^b\big) $; and
\eqref{eq:app8} follows from the expression for the capacity of the Gel'fand-Pinsker
channel \eqref{eq:cap3}, and the fact that
$ \sum_{k=1}^b  \mathbb{P}[V_k = 0] = 
  \sum_{k=1}^b  \mathbb{P}[\tau \geq k] = \mathbb{E}[\tau] \leq \ell $.

It is left to justify \eqref{eq:app4}.
For fixed-length codes (for which $ \tauhat = \ell = b $ and hence $ \prob[V_k=0] = 1 $
for all $ k < b $), the so-called Csisz\'{a}r sum identity \cite[p.\ 25]{elgamal+kim}
can be used in this step to establish \emph{equality} in \eqref{eq:app4}.
In our notation this identity has the form:
$ \sum_{k=1}^b  I\big(Y^{k-1}; S_k | M, S_{k+1}^b \big) =
  \sum_{k=1}^b  I\big(Y_k; S_{k+1}^b | M, Y^{k-1} \big) $.
It does not apply in our case due to the factors $ \prob[V_k=0] $ appearing in the
sums \eqref{eq:app31} and \eqref{eq:app4}.
However, one can establish the inequality in \eqref{eq:app4} by using the following
monotonicity property of the coefficients $ \prob[V_k=0] $, which follows directly
from the definition \eqref{eq:Vn}:
\begin{align}
\label{eq:appmon}
  j < k  \quad  \Rightarrow  \quad  \prob[V_j=0] \geq \prob[V_k=0]  .
\end{align}
We have:
\begin{align}
\nonumber
  &\sum_{k=1}^b  \prob[V_k=0]\, I\big(Y^{k-1}; S_k | M, S_{k+1}^b \big)   \\
\label{eq:app9}
    &\;=    \sum_{k=1}^b  \prob[V_k=0] \sum_{j=1}^{k-1} I\big(Y_j; S_k | M, Y^{j-1}, S_{k+1}^b \big)   \\
\label{eq:app11}
		&\;\leq \sum_{k=1}^b  \sum_{j=1}^{k-1} \prob[V_j=0]\, I\big(Y_j; S_k | M, Y^{j-1}, S_{k+1}^b \big)   \\
\label{eq:app13}
    &\;=    \sum_{j=1}^b \prob[V_j=0] \sum_{k=j+1}^{b}  I\big(Y_j; S_k | M, Y^{j-1}, S_{k+1}^b \big)   \\
\label{eq:app14}
    &\;=    \sum_{j=1}^b \prob[V_j=0]\, I\big(Y_j; S_{j+1}^b | M, Y^{j-1} \big)  ,
\end{align}
where \eqref{eq:app9} and \eqref{eq:app14} are obtained from the chain rule for
mutual information; \eqref{eq:app11} follows from \eqref{eq:appmon}; and it is
understood that $ S_{b+1}^b = Y_0 = \emptyset $, as usual.
This implies \eqref{eq:app4} and concludes the proof.
\end{IEEEproof}


\vspace{6mm}

\IEEEtriggeratref{3}

\end{document}